\documentclass{aa}
\usepackage{natbib}
\usepackage{epsfig}
\def\ut#1{\mathop{\vtop{\ialign{##\crcr
     $\hfil\displaystyle{#1}\hfil$\crcr\noalign
     {\kern1pt\nointerlineskip}\hbox{$\hfil\sim\hfil$}\crcr
     \noalign{\kern1pt}}}}}

\def\undersymbol#1#2{\mathop{\vtop{\ialign{##\crcr
     $\hfil\displaystyle{#2}\hfil$\crcr\noalign
     {\kern1pt\nointerlineskip}\hbox{$\hfil#1\hfil$}\crcr
     \noalign{\kern1pt}}}}}
\def\arcsec{^{\prime\prime}}
\def\arcmin{^{\prime}}
\def\degr{^\circ}

\usepackage{graphicx}
\usepackage{color}

\begin{document}

\title{Triangulum galaxy viewed by Planck}
       \author{F. De Paolis\inst{1,2}, V.G. Gurzadyan\inst{3,4},  A.A. Nucita\inst{1,2}, L. Chemin\inst{5}, A. Qadir\inst{6},  A.L. Kashin\inst{3}, H.G. Khachatryan\inst{3}, S. Sargsyan\inst{3}, G. Yegorian\inst{3}, G. Ingrosso\inst{1,2}, Ph. Jetzer\inst{7},   \and D. Vetrugno\inst{8}}
       
              \institute{
              Dipartimento di Matematica e  Fisica ``Ennio De Giorgi'', Universit\`a del Salento, Via per Arnesano, I-73100, Lecce, Italy  
               \and INFN, Sezione  di Lecce, Via per Arnesano, I-73100, Lecce, Italy 
               \and Center for Cosmology and Astrophysics, Alikhanian National Laboratory and Yerevan State University, Yerevan, Armenia 
              \and SIA, Sapienza University of Rome, Rome, Italy 
              \and 
             LAB, CNRS UMR 5804, Universit\'e de Bordeaux, F-33270, Floirac, France
             \and
School of Natural Sciences,
National University of Sciences and Technology, Islamabad,
Pakistan
              \and
Physik-Institut, Universit\"at
Z\"urich, Winterthurerstrasse 190, 8057 Z\"urich, Switzerland
\and 
Department of Physics, University of Trento, I-38123 Povo, Trento, Italy and 
TIFPA/INFN, I-38123 Povo,  Italy
}

   \offprints{F. De Paolis, \email{francesco.depaolis@le.infn.it}}
   \date{Submitted: XXX; Accepted: XXX}

 \abstract{We used {\it Planck} data to study the M33 galaxy and find a substantial temperature asymmetry with respect to its minor axis projected onto the sky plane. This temperature asymmetry correlates well with the HI velocity field at 21 cm, at least within a galactocentric distance of $0.5\degr$, and it is found to extend up to about $3\degr$ from the galaxy center. We conclude that the revealed effect, that is, the temperature asymmetry and its extension, implies that we detected the differential rotation of the M33 galaxy and of its extended baryonic halo. 
 }


   \authorrunning{De Paolis et al.}
   \titlerunning{Triangulum galaxy viewed by Planck}
   \maketitle
%

\section{Introduction}
A temperature asymmetry has been detected in {\it Planck} data toward the M31, Cen A, and M82 galaxies, always aligned with respect to the expected galaxy spin (\citealt{depaolis2011,depaolis2014,depaolis2015,gurzadyan2015}). The aim of this paper is to investigate whether the same phenomenon also occurs for M33 (also known as the Triangulum galaxy), which is an appropriate object on which to test our method, and on
which we can trace the dynamics of the baryonic galactic halo (which at large scales also  has to reflect the dark matter contribution) in a  model-independent way because it is  sufficiently extended and useful multi-wavelength observations are available.

M33 (or NGC 598) is the nearest late-type spiral galaxy that
lies at a distance of only about 840 kpc \citep{magrini2007,gieren2013}. It is the third largest member  (after M31 and the Milky Way) of the Local Group (see \citealt{rg} for the dynamics and substructure of the Local Group), and the  coordinates of its center are RA  $01^h 33^m 50.9^s$ and Dec $30\degr 39\arcmin 37 \arcsec$. It is classified as an SAcd galaxy, that is, a late-type spiral with a weak bar (\citealt{corbelliwalterbos2007}), no clear evidence of any bulge component (see, e.g., \citealt{bothan1992}), and relatively loosely wound arms \citep{buta2007}. At the M33 distance $1\arcmin$ corresponds to 244 pc ($1\degr\simeq 14.67$ kpc),
which allows us to study this galaxy with a great degree of accuracy. The relatively small inclination angle $i\simeq 56\degr$ \citep{vandenbergh2000} allows us to obtain a comprehensive view of the M33 galaxy, which is a key advantage, for example, for studying the correlation of the velocity field with the galaxy geometry. For a recent and comprehensive review about M33 we refer to \cite{hodge2012}. 

\begin{figure}[h!]
 \centering
  \includegraphics[width=0.46\textwidth]{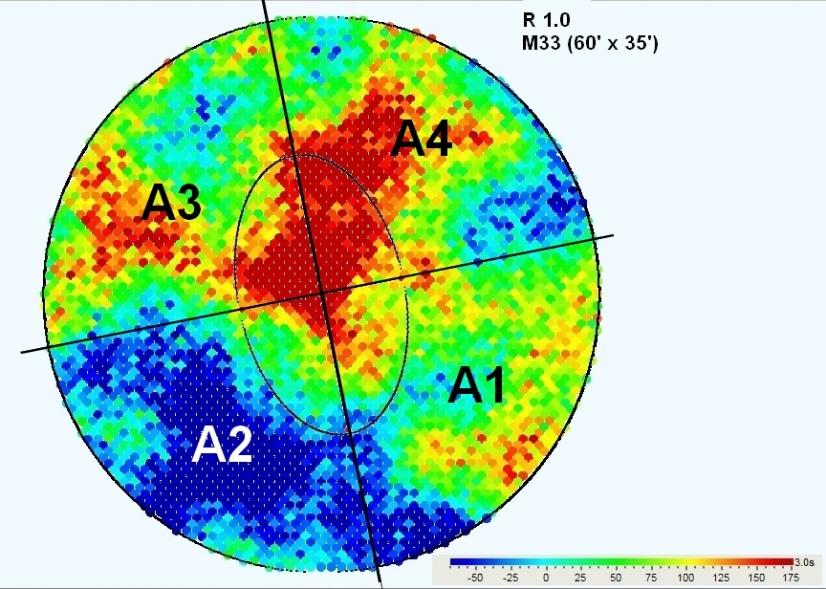}
 \caption{{\it Planck} field toward the M33  galaxy in the 70 GHz band. The pixel color gives the temperature excess  in $\mu$K with respect to the mean CMB temperature. The optical extension of the M33 galaxy is indicated by the inner ellipse with major  and minor axes of  $60\arcmin$ and $35\arcmin$, respectively. The four quadrants A1, A2, A3, and A4 are used in our analysis.} \label{fig1}
 \end{figure}
The M33 disk is rotating with a maximum circular velocity of about  $120-130$ km s$^{-1}$ and has a rising profile  (see, e.g., \citealt{corbellisalucci2000,corbelli2014,kam2015}), indicating that the outer regions of the galaxy possess
substantial mass. The observations at wavelength of 21 cm show the detailed spatial and kinematic structure of the neutral hydrogen
in M33, which extends farther out than the stellar component. Interestingly, at least $18\%$ of the HI gas is found beyond the star-forming disk.  There is some evidence of a faint  halo component (\citealt{mcconnachie2010,cockcroft2013}) that was previously undetected (see, e.g., \citealt{trinchieri1988,barker2007}). The study in the microwaves of the M33 disk and halo within $4\degr$ is the aim of the following sections, particularly with the aim of investigating their rotation.

\section{Planck data analysis and results}

\begin{figure}[h!]
 \centering
  \includegraphics[width=0.45\textwidth]{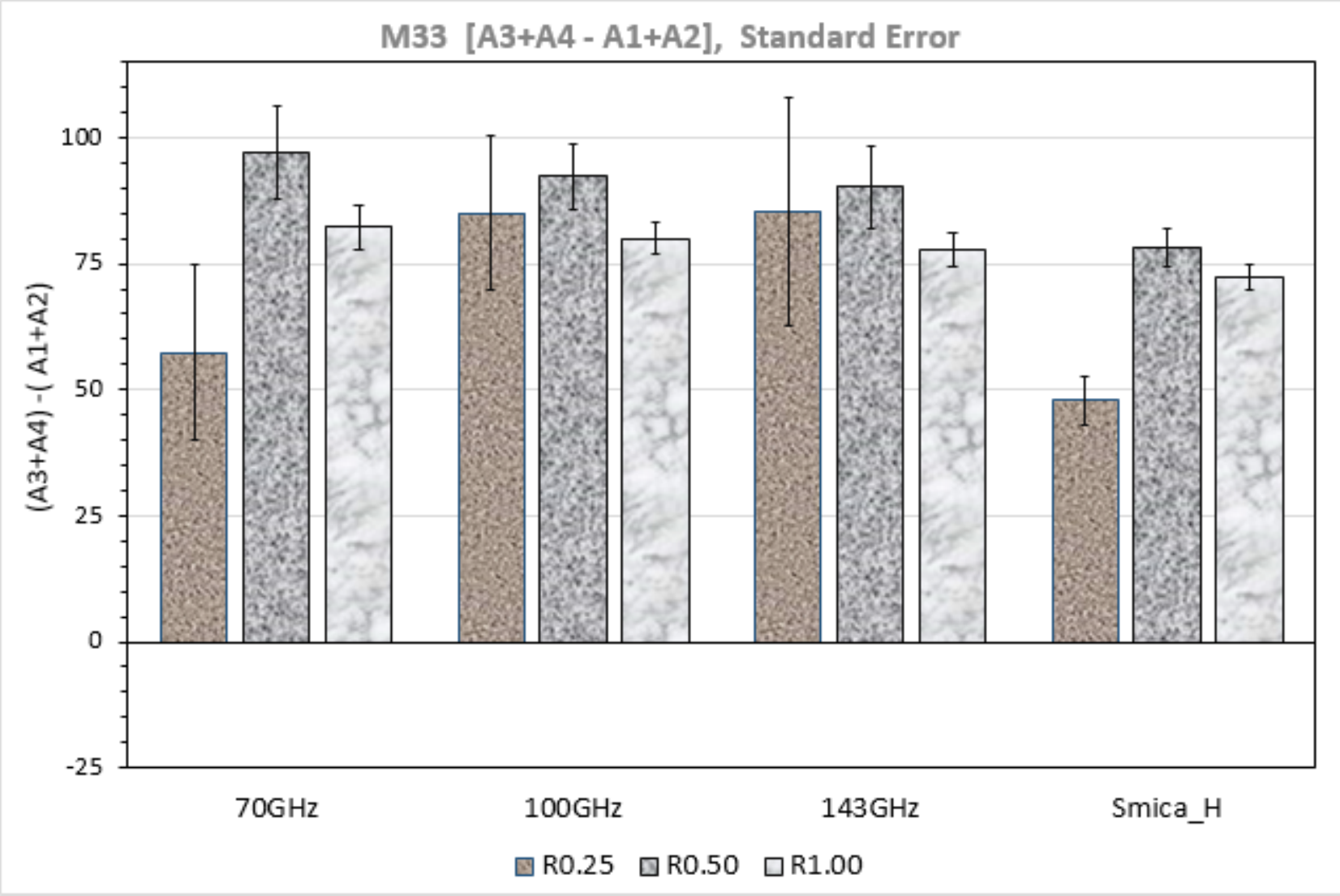}
    \includegraphics[width=0.45\textwidth]{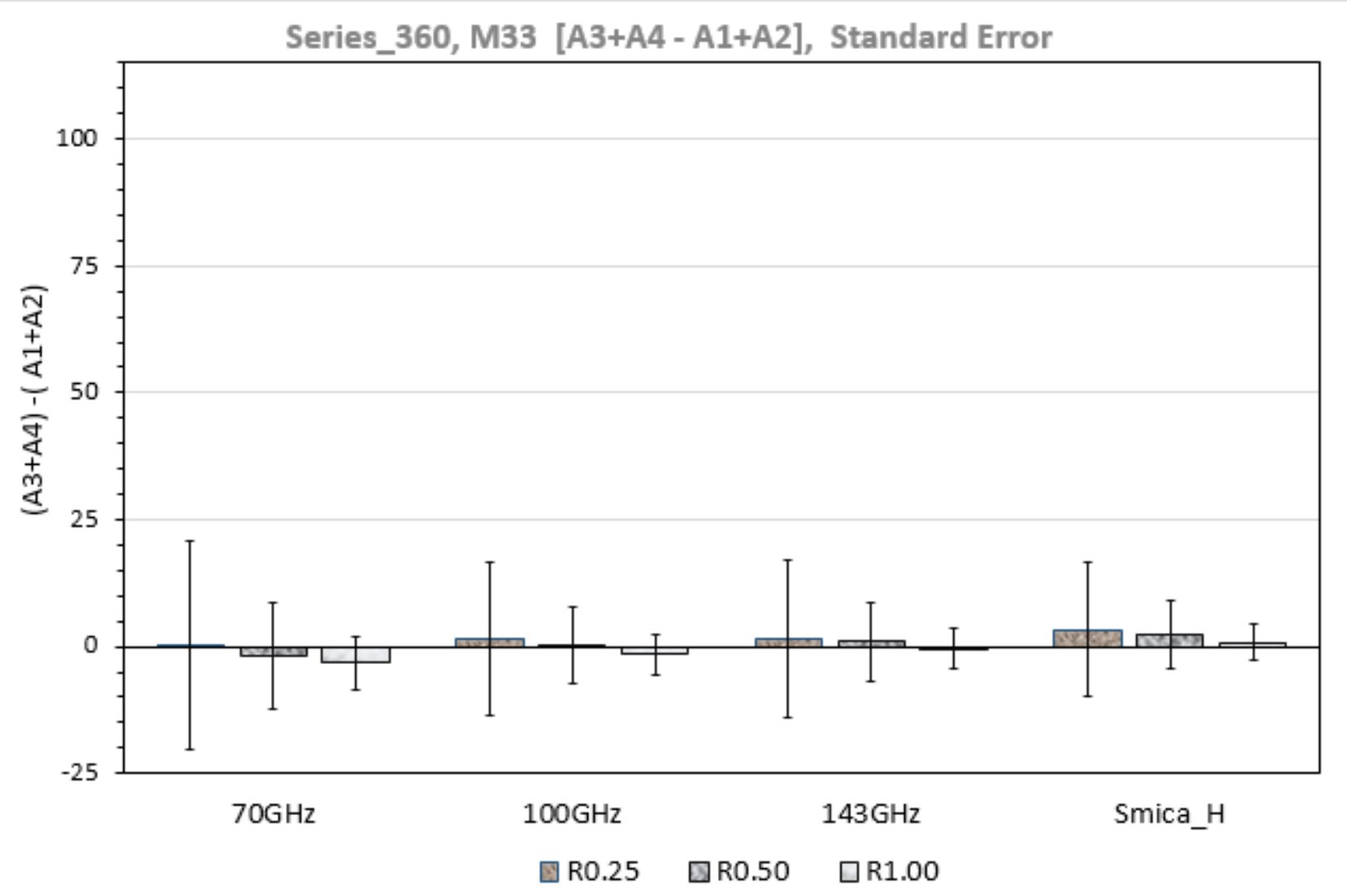}
  \includegraphics[width=0.52\textwidth]{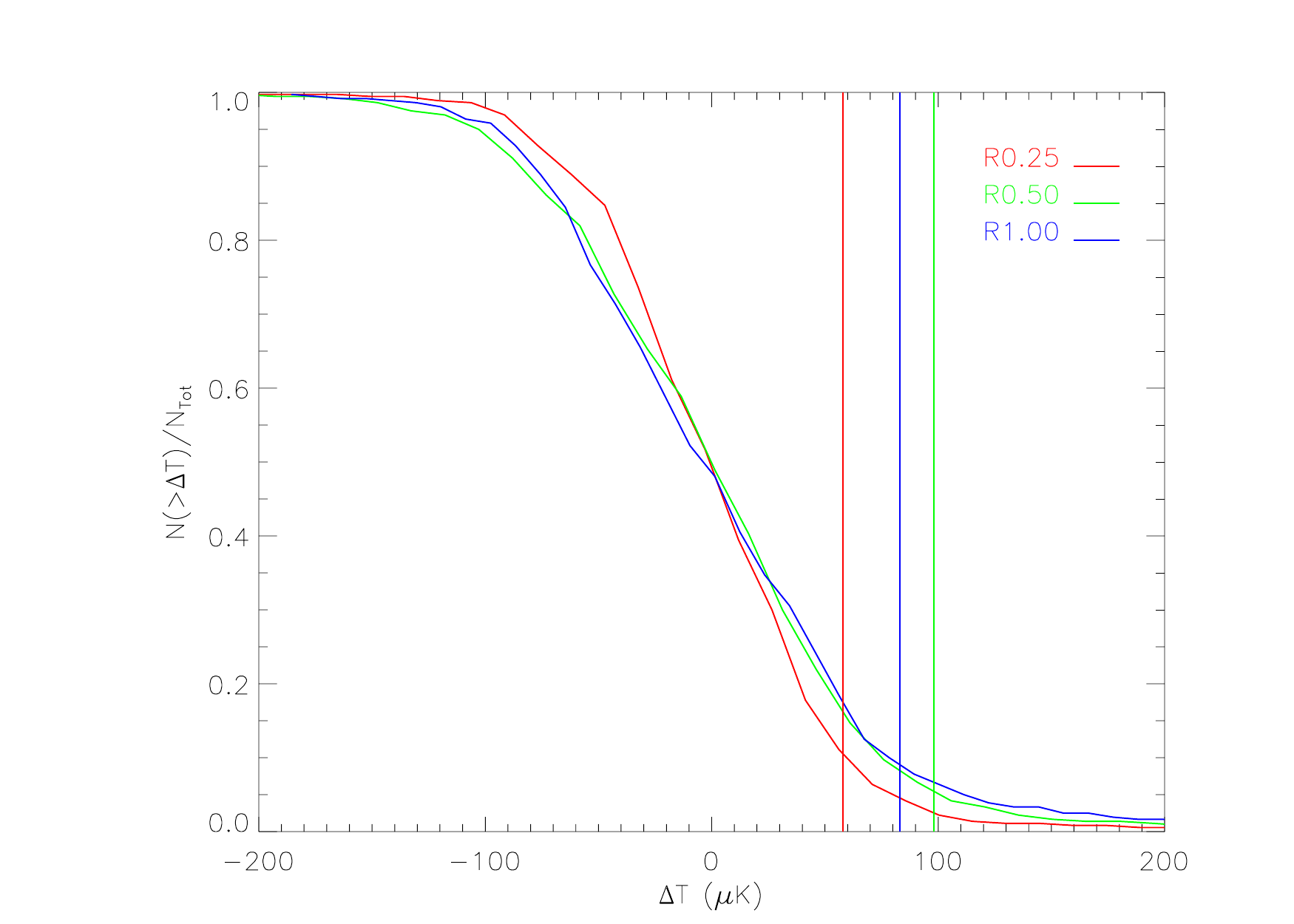}
 \caption{Upper panel: the temperature asymmetry toward M33 in $\mu$K (with the standard errors) of regions A3 and A4 with respect to regions A1 and A2 in the  {\it Planck} bands within three radial distances of  $15\arcmin$ ($R0.25$), $30\arcmin$ ($R0.50$), and $1\degr$ ($R1.00$). Central panel: the same for the 360 control fields with the same geometry (shown in Fig. \ref{fig1}) equally spaced at one degree distance to each other in Galactic longitude and at the same latitude as M33. We also present (indicated with SMICA\_H) the result obtained from the foreground-cleaned map. Bottom panel:  cumulative distribution of the temperature asymmetry values in the 70 GHz {\it Planck} band within $R0.25$, $R0.50,$ and $R1.00$. The vertical lines show for each region the value measured toward M33.}
 \label{fig2}
 \end{figure}

We used data of {\it Planck} 2015 release \citep{planck2015a} in the bands at 70 GHz of the Low Frequency Instrument (LFI), and in the bands at 100 and 143 GHz of  the High Frequency Instrument (HFI) with angular resolution of $13\arcmin$, $9.6\arcmin$, and $7.1\arcmin$ in terms of FHWM, respectively. The frequency maps  are provided in cosmic microwave background (CMB) temperature at resolution corresponding to $N_{side}$=2048 in HEALPix scheme \citep{gorski2005,planck2015b}.
To begin,  the considered region of the sky (shown in Fig. \ref{fig1} in the {\it Planck} band at 70 GHz) toward the M33 galaxy 
was divided into  four quadrants indicated as A1, A2, A3, and  A4.
The optical extension of M33 of $ 60\arcmin \times 35 \arcmin$ (\citealt{berkhuijsen1983}) is also shown in the same figure.
The mean
temperature excess in $\mu$K (i.e., the average pixel  temperature excess with respect to the mean CMB temperature)  in each of the indicated 
regions was obtained in the  {\it Planck} bands at 70, 100, and 143 GHz with
the corresponding standard error. The standard error has been 
calculated as the standard deviation of the excess temperature
distribution divided by the square root of the pixel number in
each region. We have verified that within the errors, the sigma
values calculated in this way are consistent with those evaluated
by using the covariance matrix obtained by a best-fitting
procedure with a Gaussian to the same distribution. The obtained results are presented in the upper panel of Fig. \ref{fig2}, which show the temperature  asymmetry of regions A3 and A4 with respect to regions A1 and A2  within $15\arcmin$, $30\arcmin$, and $1\degr$  from the center of M33.
This together with Fig. \ref{fig1} shows that regions A3  and
A4 are systematically hotter than regions A1 and A2  in the  considered {\it Planck} bands. The detected temperature asymmetry $\Delta T$ with respect to the minor axis of the M33 galaxy is clearly  aligned along its rotational direction,  indicating a Doppler-induced effect modulated by the spin of the galaxy, since it is almost independent of the {\it Planck} band. This result is also confirmed by  analyzing the foreground-corrected SMICA map   \citep{planck2015a}, as indicated by the histogram at the right side of the upper panel in Fig. \ref{fig2}.
In this respect, we also note that  the temperature asymmetry  in the SMICA map does not mean that it is of cosmological origin. To remove the CMB contribution from a {\it Planck} map,  it is therefore not possible to  subtract the SMICA map on a pixel-by-pixel  basis from the considered
data (see, e.g., Appendix B in \citealt{hermelo2016}).

 In the central panel of Fig. \ref{fig2} we show the temperature asymmetry for 360 control fields with the same geometry as in Fig. \ref{fig1}, equally spaced at one degree distance to each other in Galactic longitude and at the same latitude as M33, together with the corresponding standard error of the mean.  To test more carefully whether the measured values of $\Delta T$ towards M33 is real or may be a fluctuation of the CMB background, we show in the bottom panel of Fig. \ref{fig2} the cumulative distribution (i.e., the relative number of control fields with a stronger temperature asymmetry than a given $\Delta T$) of the temperature asymmetry in the 70 GHz {\it Planck} band. The vertical lines show for each region the value measured toward M33. Fewer than $5-10\%$ of the control fields have $\Delta T$ values higher than the corresponding values toward M33, and moreover, a careful inspection of the single profiles shows that only eight of the control field profiles are globally above the profile
of M33, implying  a probability certainly  $\ut <2.2\%$ that the detected temperature asymmetry toward the M33 galaxy is due to a random fluctuation of the CMB sky (see, in addition, the discussion at the end of Sect. 3, where we consider the region up to $4\degr$ from the center of M33).

\begin{figure}[h!]
 \centering
  \includegraphics[width=0.46\textwidth]{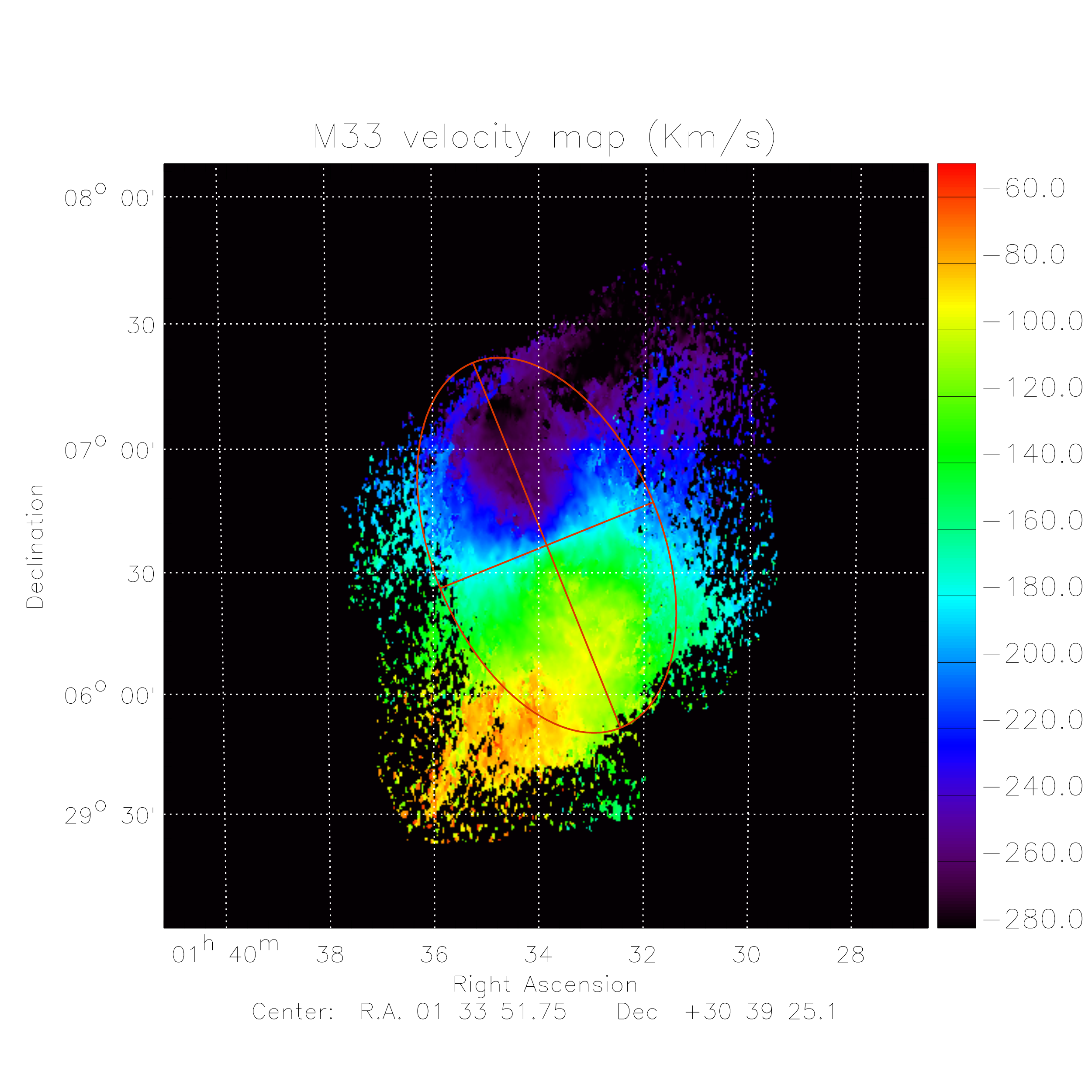}
 \caption{HI velocity field (in km s$^{-1}$) at 21 cm of  the M33 galaxy (from \citealt{chemin2012}). The color of each pixel shows the line-of-sight velocity with the color scale indicated in the right part of the panel. The M33 ellipse with a semi-major  axis equal to $48\arcmin$ and a semi-minor axis of  $28\arcmin$ ($60\%$ larger than that in Fig. \ref{fig1}) is also shown.} \label{fig3}
 \end{figure}
  \begin{figure}[h!]
 \centering
   \includegraphics[width=0.38\textwidth]{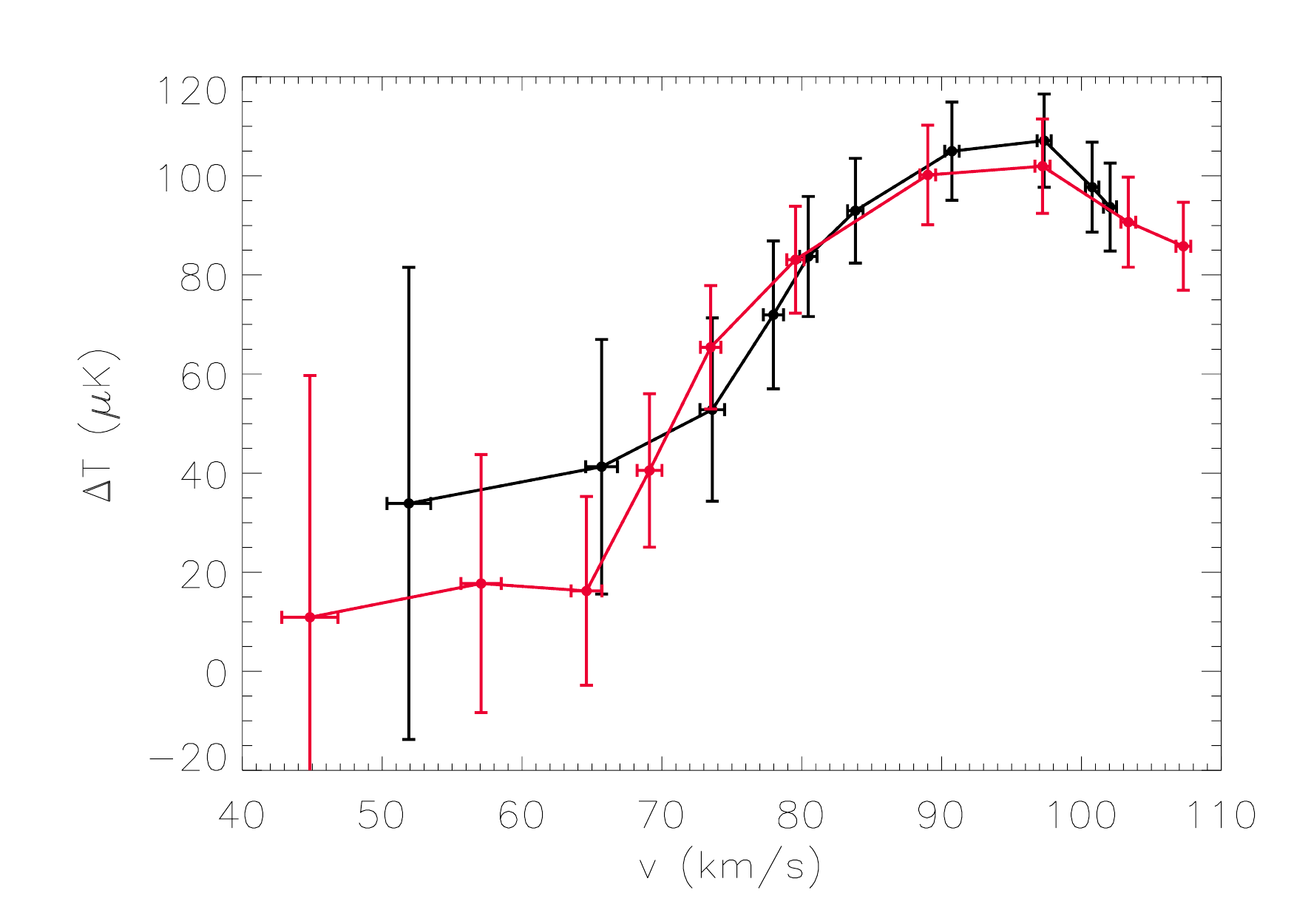}
  \includegraphics[width=0.36\textwidth]{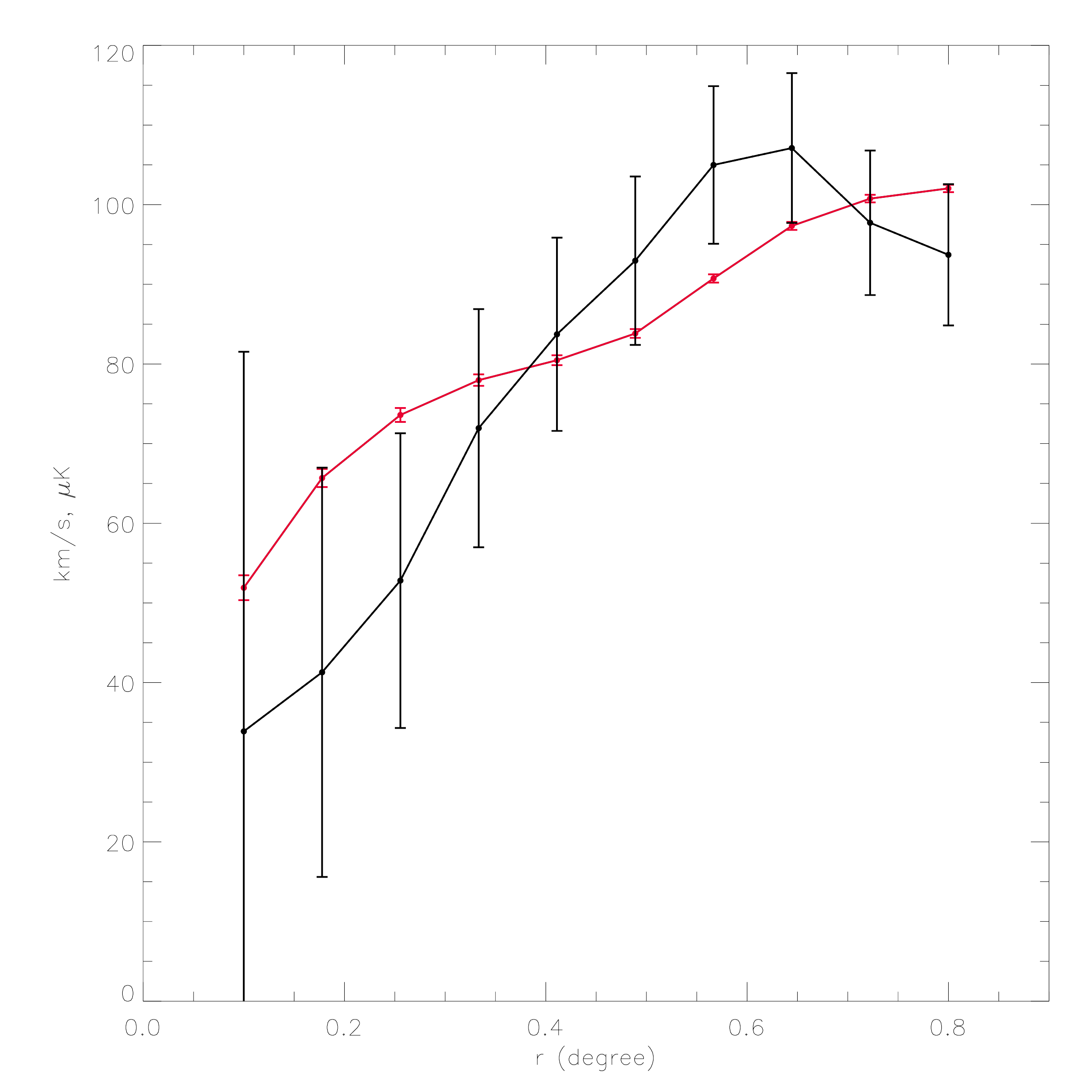}
    \includegraphics[width=0.36\textwidth]{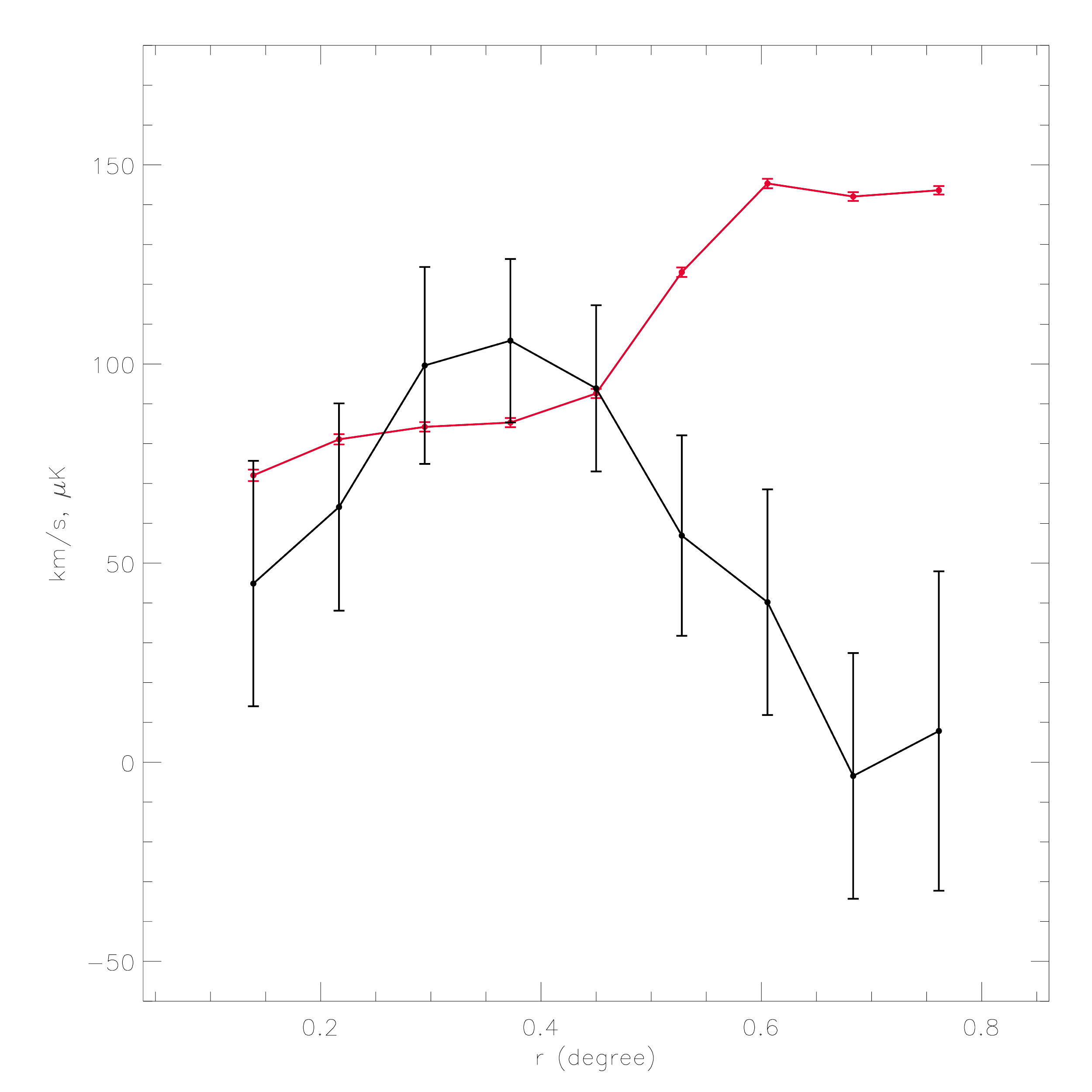}
 \caption{Upper panel: correlation of the temperature asymmetry with respect to the 21 cm velocity field within ten ellipses with major axes between $0.1\degr$ and $0.8\degr$ (black line) and for the ellipse with position angle rotated by $34\degr$ (red line). Central panel: radial profile, in $\mu$K for the temperature asymmetry  in the 70 GHz {\it Planck} band (black curve) and for the 21 cm DRAO data relative to the M33 HI velocity field (red profile, 
 in km s$^{-1}$) in Fig. \ref{fig3}. Bottom panel: temperature and velocity  radial profiles, but with the temperature and velocity asymmetry calculated shell by shell. }
 \label{fig4}
 \end{figure}
  \begin{figure}[h!]
 \centering
   \includegraphics[width=0.43\textwidth]{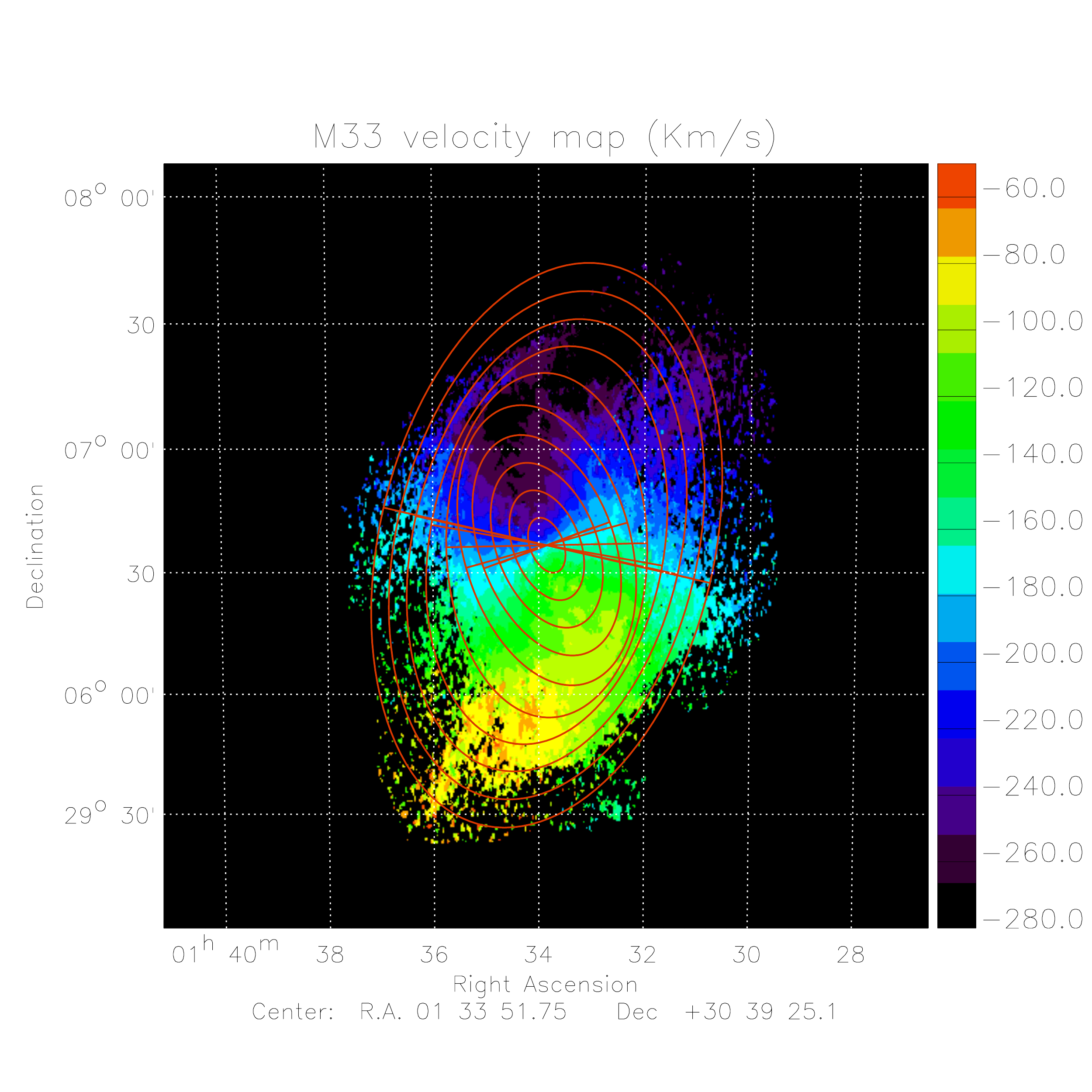}
  \includegraphics[width=0.43\textwidth]{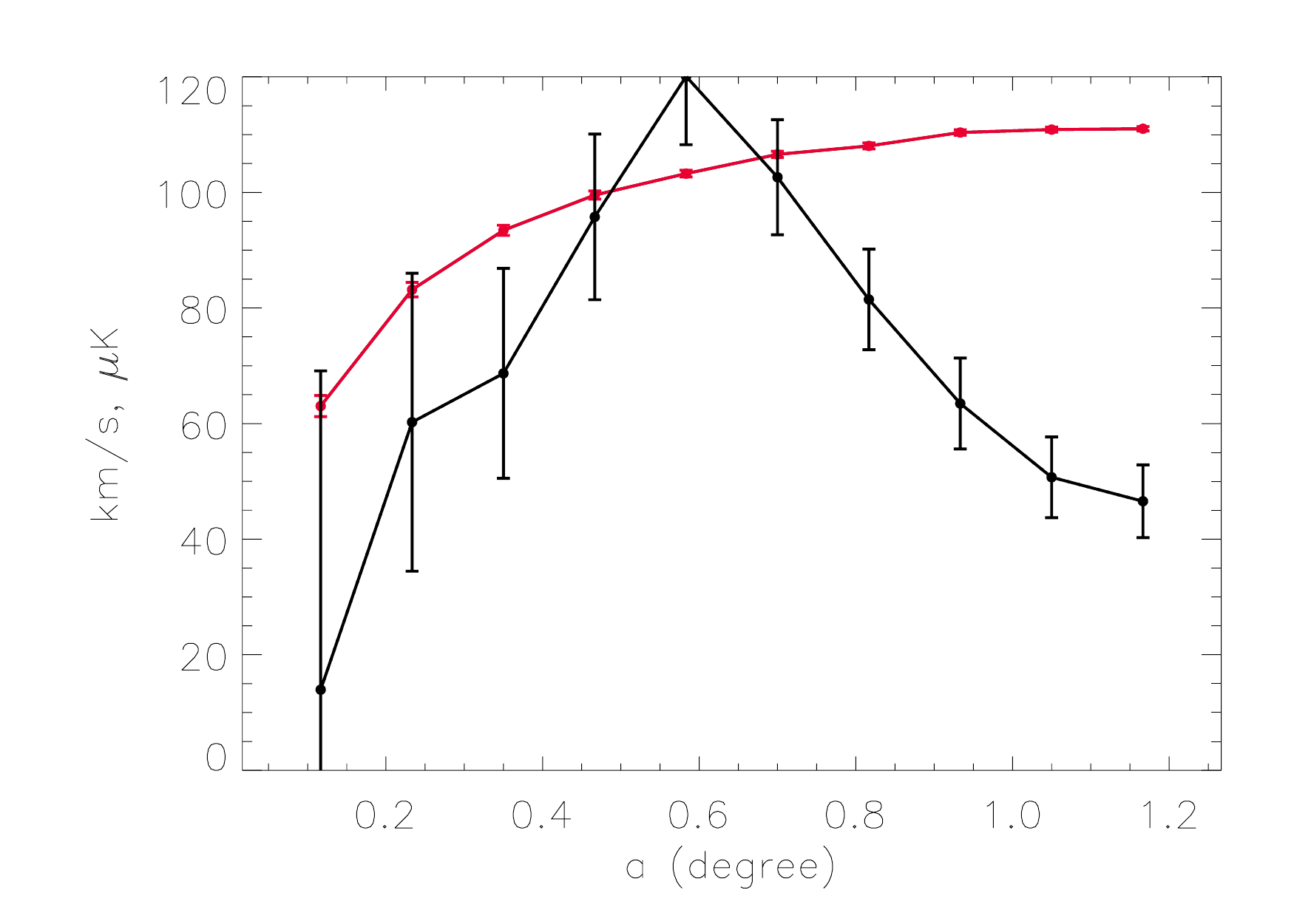}
    \includegraphics[width=0.43\textwidth]{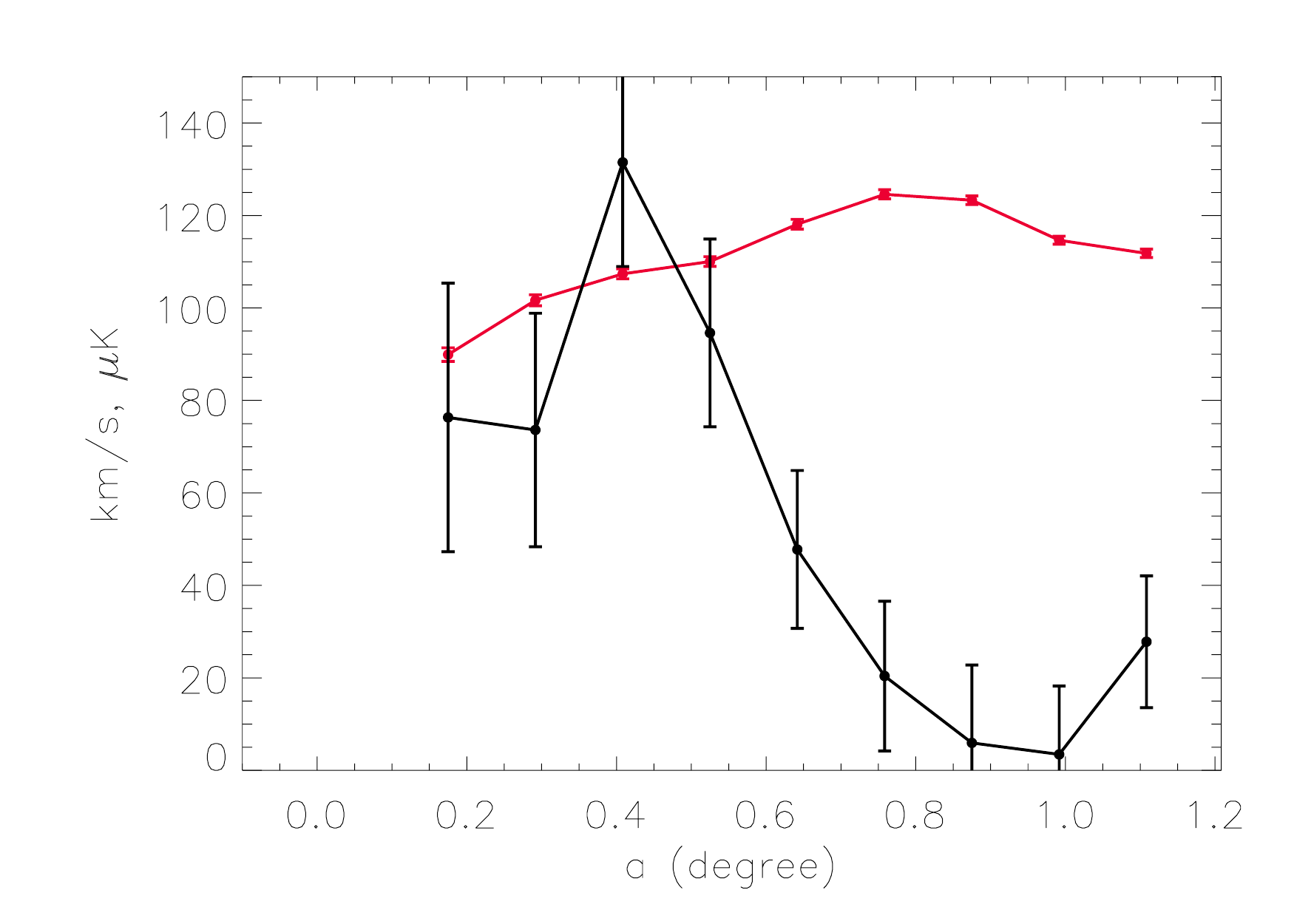}
 \caption{Upper panel: the considered ellipse geometry from the model in \cite{corbellischneider1997} superposed on the HI velocity field  at 21 cm of  the M33 galaxy (from \citealt{chemin2012}). For each ellipse we also show the minor axis with respect to which the temperature asymmetry and HI velocity are  calculated. Median panel: radial profile, in $\mu$K, for the temperature asymmetry  in the 70 GHz {\it Planck} band (black curve) and for the 21 cm DRAO data relative to the M33 HI velocity field (red profile). The points in the horizontal axis show the central value in degrees of the semi-major axes of two adjacent ellipses (i.e., $(a_i+a_{i+1})/2$).  Bottom panel: temperature and velocity  radial profiles, but with the temperature and velocity asymmetry calculated shell by shell. }
 \label{figcorbelli}
 \end{figure}

\section{Comparison with 21 cm data}
The M33 galaxy has been studied in great detail in the 21 cm emission line of the neutral hydrogen (see, e.g., \citealt{corbellisalucci2000,chemin2012,keenan2016}), and the obtained velocity maps clearly show the  M33 rotation with respect to its minor axis. The same effect is also clear in other analyses,  such as the CO velocity field (Fig. 6 in \citealt{gratier2010}). The comparison  of {\it Planck} data  with  the HI velocity field at 21 cm might in principle helpful to infer the possible source of the temperature asymmetry found in {\it Planck} data. 
If the signal in the microwave band were due to gas or dust  emission, modulated by the M33 spin, the temperature asymmetry between regions A3 and A4 and A1 and A2 should be given by
 \begin{equation}
 \Delta T/T=2\frac{v\sin i}{c}S\langle\tau\rangle \label{eq1} 
 \end{equation}
 according  to the model
first proposed in \cite{depaolis1995}. Here, $v$ is the rotation velocity, $i\simeq 56\degr$ is the M33 disk inclination angle, $S$ is the gas or dust filling factor, and $\langle\tau\rangle$ is the averaged optical depth within a given {\it Planck} band. The shape of the temperature and velocity asymmetry profiles should then correlate tightly, provided that $S\langle\tau\rangle$ does not change much within the galaxy.
To investigate this question, we considered the  velocity field reported in \cite{chemin2012} that was acquired with  aperture synthesis observations at the Dominion Radio Astrophysical Observatory (DRAO). In Fig. \ref{fig3} we show the velocity field of the HI component with the superimposed M33 ellipse (red curve). We
note that the ellipse shown in Fig. \ref{fig3} and considered in the following analysis is slightly larger than the ellipse shown in Fig. \ref{fig1} (which corresponds to the optical extension of the M33 galaxy).  The semi-major axis was $30\arcmin$ in Fig. \ref{fig1}, while it is  $0.8\degr=48\arcmin$ in Fig. \ref{fig3}. The reason for this choice is that we wished to compare {\it Planck} and DRAO data at the largest possible scale (limited by the extension of the available  21 cm data), which becomes clear in the following discussion.

In the upper panel of Fig. \ref{fig4} we present the temperature asymmetry in the 70 GHz {\it Planck} band versus the M33 rotation velocity evaluated within ten ellipses with semi-major axes between $0.1\degr$ and $0.8\degr$ (black line). However, the M33 HI disk is known to have a prominent warp, which is mostly consistent with a  twist of the kinematic position angle as the inclination remains more or less constant with radius (\citealt{corbelli2014}). From a galactocentric distance of about 7 kpc, the kinematic position angle of the major axis of the outer regions is seen to decline by about $34\degr$ with respect to the inner disk regions.
To investigate whether the $\Delta T-v$ correlation is modified by the warping HI disk,  we also considered a different ellipse with a position angle rotated by $34\degr$  (with respect to that shown in Fig. \ref{fig3}) and obtained the red curve in the upper panel of Fig. \ref{fig4}. 
 \begin{figure}[h!]
 \centering
  \includegraphics[width=0.42\textwidth]{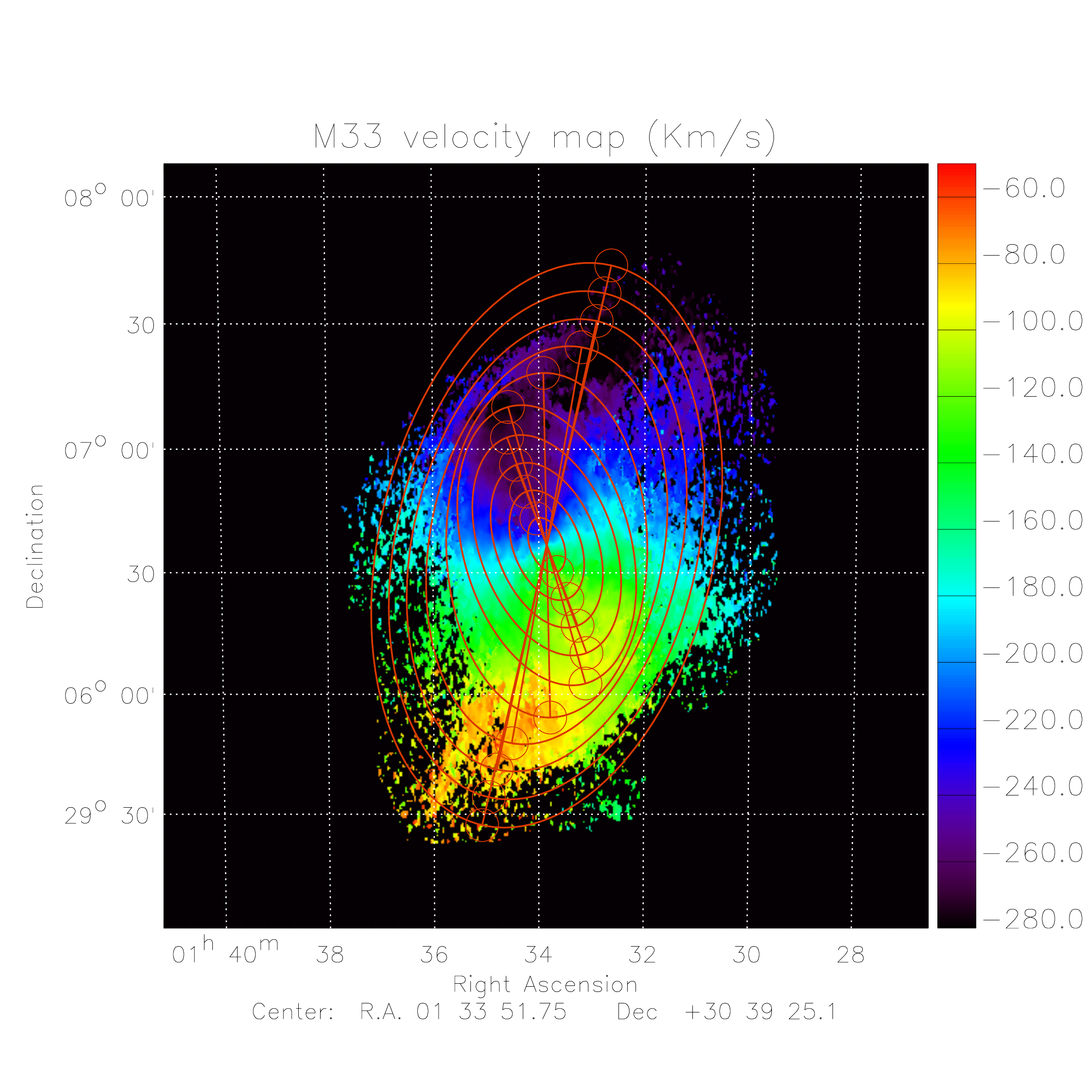}
  \includegraphics[width=0.50\textwidth]{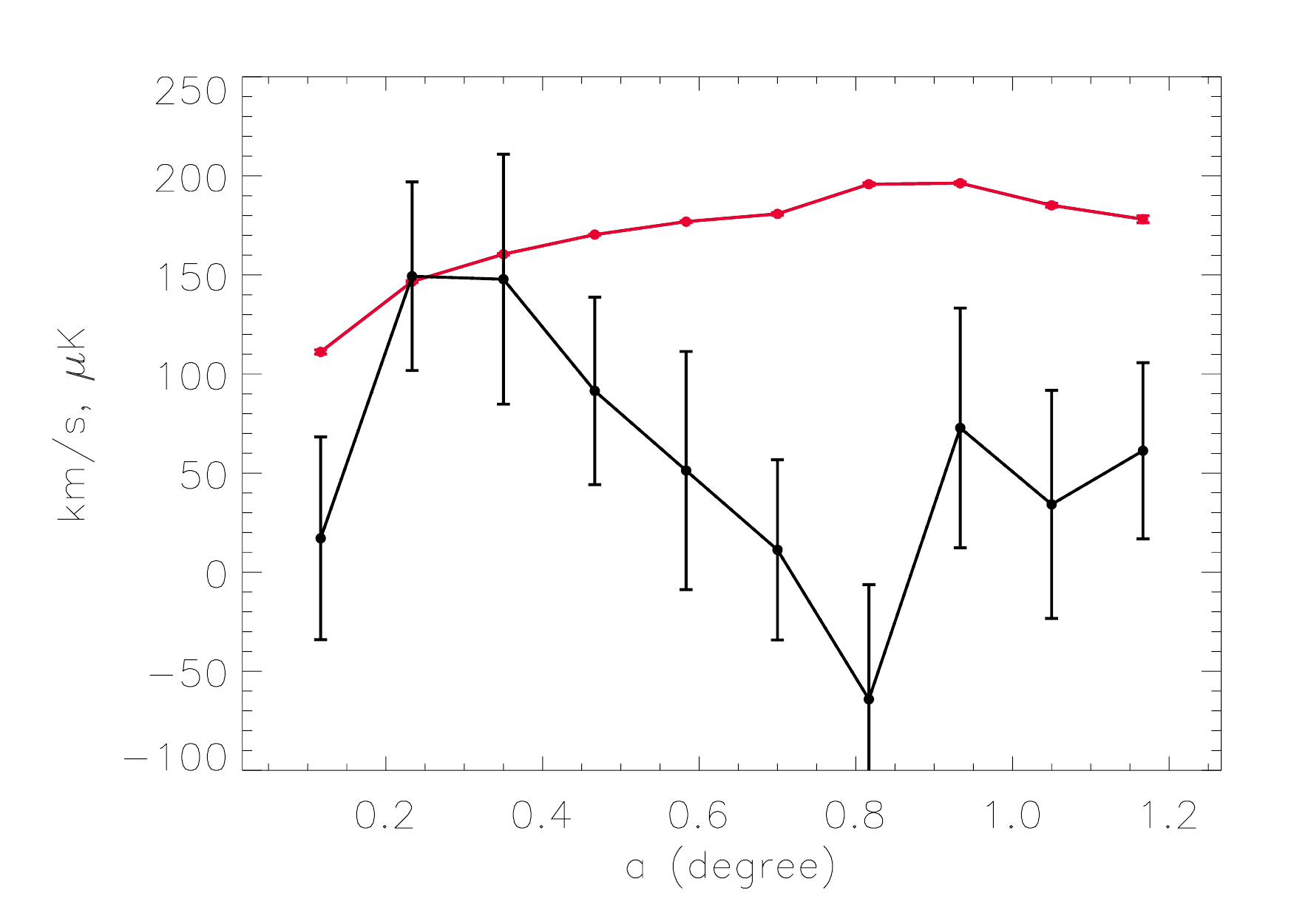}
 \caption{Upper panel: the M33 HI velocity field with the position of $4\arcmin$ circles used to estimate (bottom panel) the velocity (red curve) and temperature (in the 70 GHz {\it Planck} band; black curve)  asymmetry profiles, as described in the text.} 
 \label{corbelli_circles}
 \end{figure}
\begin{figure}[h!]
 \centering
  \includegraphics[width=0.42\textwidth]{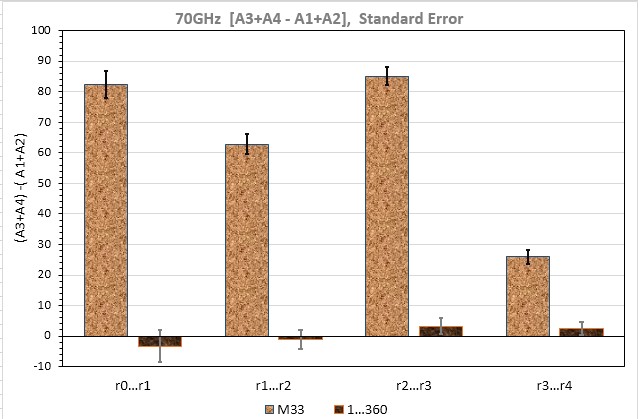}
  \includegraphics[width=0.50\textwidth]{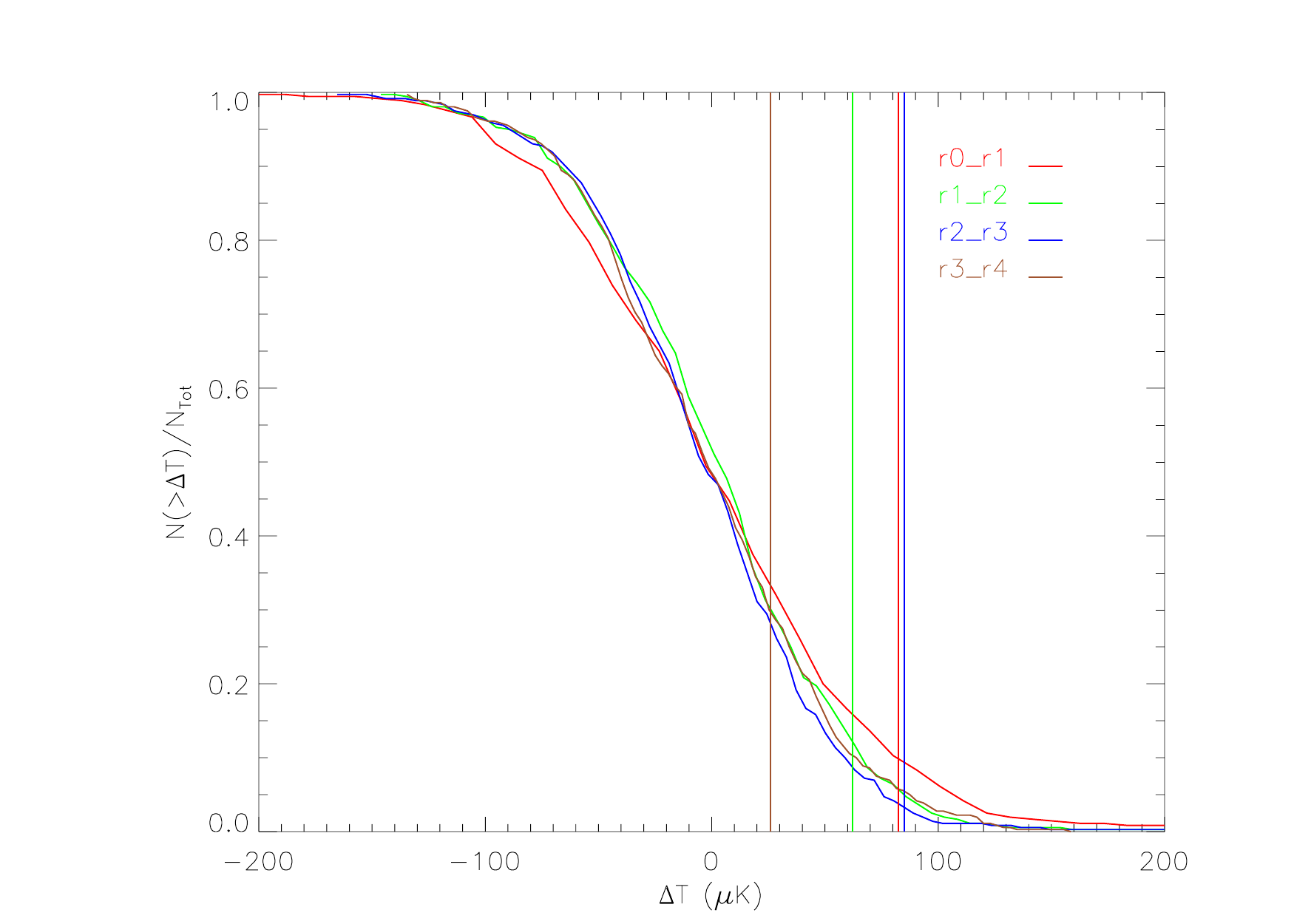}
 \caption{Upper panel: the temperature asymmetry in $\mu$K, in the 70 GHz {\it Planck} band, of regions A3 and A4 compared with
regions A1 and A2 in four annuli around M33 (light brown histograms). The dark brown histograms depict the mean $\Delta T$ value of the 360 control fields with the corresponding standard error of the mean.  Bottom panel: cumulative distribution of the 360 control field temperature asymmetry in the four regions  in the upper panel. The vertical lines show the temperature asymmetry value toward M33 for each considered region. } 
 \label{fig6}
 \end{figure}

The correlation between $\Delta T$ and $v$ for the two profiles is quite similar, and the Pearson correlation coefficient is 0.93 and 0.89 for the black and red profiles, respectively. 

In Fig. \ref{fig4} we also show
the temperature asymmetry  in the 70 GHz {\it Planck} band (black curve) and  the velocity asymmetry (red curve) radial profiles, with the respective errors. 
In particular, in the middle panel we consider ten concentric ellipses with semi-major axes between $0.1\degr$ and $0.8\degr$ and, within the $i_{\rm th}$ ellipese,   the average temperature  (in {\it Planck}'s 70 GHz band) and the radial velocity asymmetry (with respect to the galaxy minor axis) are calculated.  The agreement between the temperature and velocity curves is quite good, although the errors relative to the temperature asymmetry curve are quite large. 
In the bottom panel of Fig. \ref{fig4} we present the radial profiles but, different from the middle panel,  with the temperature and velocity asymmetry calculated shell by shell (in a symmetric way with respect to the minor axis of the M33 galaxy). The two profiles  apparently agree well within about $0.5\degr$,  the extension of the visible M33 disk, while the agreement seems lost beyond $0.5\degr$. 

A more detailed study of this problem is presented in Fig. \ref{figcorbelli}, where we consider a set of ten ellipses with increasing semi-major axis values  and position angles rotated from  $21.1\degr$ (corresponding to the ellipse directed as in Figs. \ref{fig1} and \ref{fig3}) to $-13 \degr$ (corresponding to the outermost  ellipse in the upper panel of Fig. \ref{figcorbelli}), following the geometrical model described in Sect. 4 in  \cite{corbellischneider1997}. The  ellipses have semi-major axes since $7\arcmin$ to $1.2\degr$ and position angle given by $PA(a)=21.1\degr -17\degr (1+\tanh [(a-40.8\arcmin)/7\arcmin]$   ($a$ being the semi-major axis, in arcmin,  of the $i_{\rm th}$ ellipse) and describe with a certain degree of accuracy the warped disk of the M33 galaxy. The obtained ellipses are  shown in the upper panel of Fig. \ref{figcorbelli}, and the corresponding temperature asymmetry and HI velocity profiles are shown in the middle and bottom panels of Fig.  \ref{figcorbelli}, which are analogous for the rotated ellipses to the corresponding panels in Fig. \ref{fig4}.
As indicated by the middle and bottom panels of Fig.  \ref{figcorbelli}, the two profiles deviate from each other beyond about $0.6\degr$, so that we can conclude that beyond this distance, the agreement of the two considered profiles is not substantially improved by the consideration of the \cite{corbellischneider1997} geometrical model with respect to the model used in Fig. \ref{fig4}.

However, the lack of agreement between the temperature asymmetry and velocity profiles beyond about $0.5\degr$ is expected. While the HI velocity field precisely yields the gas velocity in a given direction of sight, the temperature in a given {\it Planck} pixel in a certain band does depend on a number of physical parameters
that are difficult to quantify, therefore it depends on a complex superpositions of effects.  Roughly speaking,  it indicates that in the range   $0.5\degr - 1.2 \degr$ the value of  $S\langle\tau\rangle$ in Eq. (\ref{eq1}) decreases rapidly. 

We also note that the emission peak of about $120-130$ $\mu$k at about $0.5\degr$ from the center of M33 (central and bottom panels in Fig. \ref{figcorbelli}) may be related to the strong increase of the heating efficiency that rises from about $0.8\%$ in the inner $4.5$ kpc to $\simeq 3\%$ at a radial distance of about $6$ kpc \citep{kramer2013}. 

The analysis discussed above is also confirmed by Fig. \ref{corbelli_circles}, in which we have considered the temperature and velocity asymmetry between symmetrically disposed circles of $4\arcmin$ radius positioned at the extrema of each ellipse in the \cite{corbellischneider1997}  model, as indicated in the upper panel of the figure. Here we show the major axis of each ellipse at the vertex of which the $4\arcmin$ radius circles are shown. In the bottom panel of Fig. \ref{corbelli_circles} we plot the velocity (red curve) and temperature (black curve) asymmetry profiles obtained by subtracting  the average velocity and temperature values in each symmetrically positioned $4\arcmin$ circle with respect to the center of M33.  This plot traces the disk rotation
of M33 better than Fig. \ref{figcorbelli}. 
The choice of circles of this size arises from a compromise between the necessity to have a sufficient number of pixels (to have lower error bars in the temperature profile) and being at the same time representative for each step along the ellipse major axes. In spite of the relatively large error bars in the temperature asymmetry curve, the obtained profiles are required to maximize the asymmetry effect with respect to the results in Figs. \ref{fig4} and \ref{figcorbelli}. From comparing the result with that in the bottom panel of Fig. \ref{figcorbelli}, the agreement between the velocity and temperature profiles becomes slightly better, although the two curves diverge beyond about $0.4\degr$. It is also confirmed, however, that the temperature profile starts to rise again beyond  about $0.8\degr$, and the agreement between the two profiles  is restored within $2\sigma$.

We also analyzed the temperature asymmetry farther away from the center of M33. In the upper panel of Fig. \ref{fig6} we present a histogram with the analysis of {\it Planck} data in the 70 GHz band up to $4\degr$ from the M33 center. The light brown columns give the temperature asymmetry in $\mu$K of regions A3
and A4 with respect to regions A1 and A2 within $1\degr$ (indicated by $r0\_r1$) and in regions $1\degr-2\degr$  (indicated by $r1\_r2$), $2\degr-3\degr$ (indicated by $r2\_r3$),  and $3\degr- 4\degr$ (indicated by $r3\_r4$), with the corresponding asymmetry in the 360 control fields (in dark brown). The temperature asymmetry extends up to about $3\degr$ from the M33 center  and then rapidly decreases.  In the bottom panel of Fig. \ref{fig6} we show the cumulative distribution of the measured $\Delta T$ values for the 360 control fields together with the corresponding values detected toward M33 (vertical lines). With the exception of the outermost ring between $3\degr$ and $4\degr$, fewer than $5-10\%$ of the control fields have $\Delta T$ higher than the corresponding values toward M33. This corresponds to a probability $\simeq 2\times 10^{-4}$ (i.e., $0.09 \times 0.12 \times  0.05 \times 0.33$) for a random nature of the M33 temperature asymmetry.  This result is also confirmed by a direct inspection of the temperature asymmetry profiles in each of the control fields, which shows that only one of the control fields has $\Delta T$  higher than that of M33  in each of the four considered regions. This  indicates that the probability that the M33 temperature asymmetry is due to a random fluctuation of the microwave signal is $\ut < 7\times 10^{-4}$. These two considerations indicate that the temperature asymmetry observed up to about $3\degr$ from the center of M33 is a genuine effect.

\begin{figure}[h!]
 \centering
  \includegraphics[width=0.48\textwidth]{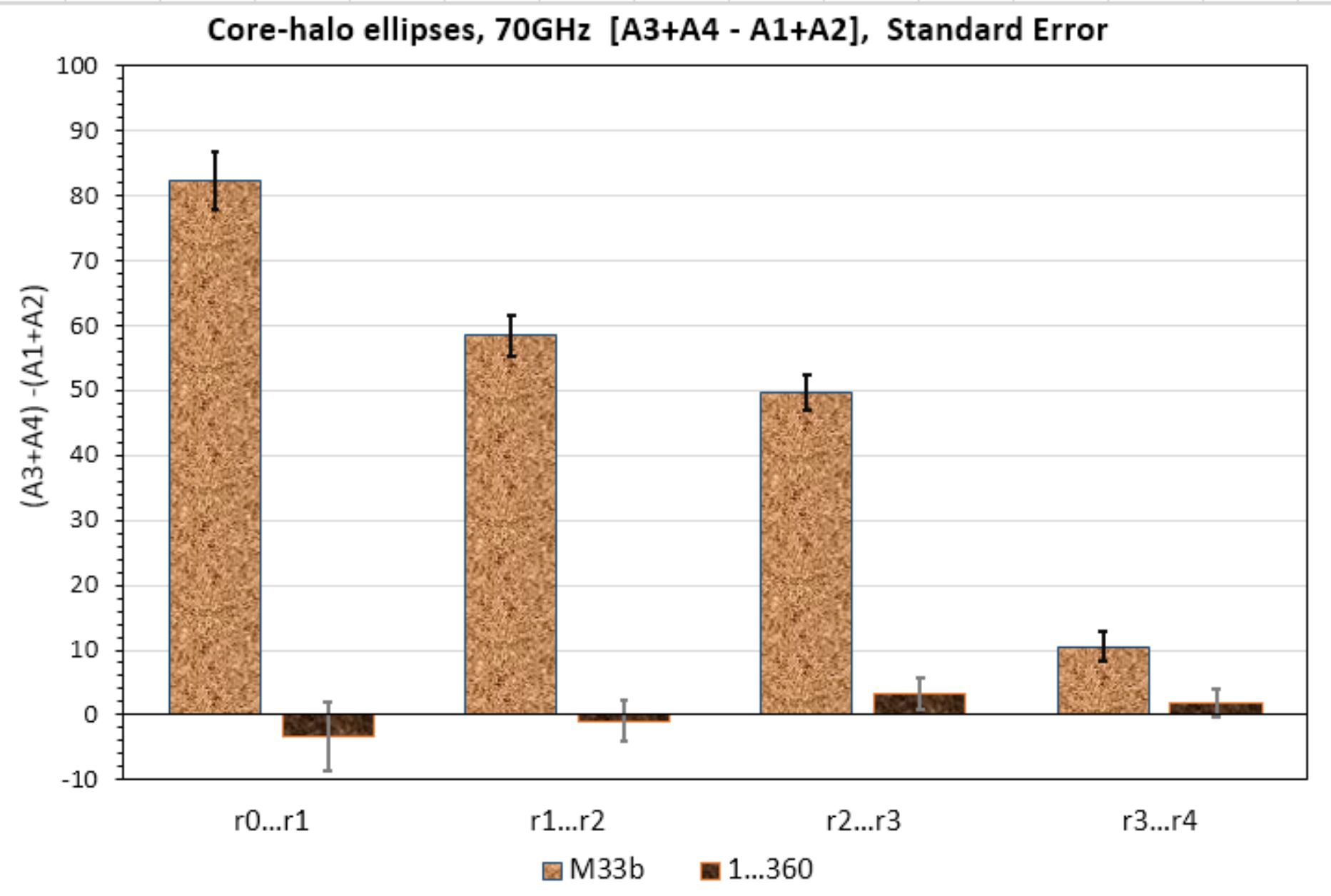}
 \caption{Temperature asymmetry in $\mu$K, in the 70 GHz {\it Planck} band,  of region A3 and A4 with respect to A1 and A2 for the second, third, and fourth annuli around M33 with PA$=-13\degr$.} 
 \label{fig7}
 \end{figure}

Finally, we  considered the same geometry as in Fig. \ref{fig6}, but with the  position angle of the second, third, and fourth annuli rotated by $34\degr$, that is, oriented as the outermost ellipse of the model by  \cite{corbellischneider1997}. In Fig. \ref{fig7} we present the obtained temperature asymmetry  in the four annuli up to $4\degr$. A comparison between the upper panel of Fig. \ref{fig6} and Fig. \ref{fig7} clearly shows that  the temperature asymmetry shows higher values for the original orientation of the geometry.

\section{Conclusions}
Similar to other nearby galaxies analyzed in some previous papers, M33 also shows a temperature asymmetry in {\it Planck} data with respect to the M33 minor axis projected onto the sky plane. This effect is clearly visible in all considered {\it Planck} bands at 70, 100, and 143 GHz,  as well as in the best foreground-corrected SMICA map. As shown by the good correlation with the HI velocity field within about $0.5\degr$, this is likely due to gas or dust emission in the M33 disk, modulated by its rotation.  After peaking  in the 70 GHz band at about $0.5\degr$,  the microwave signal decreases between $0.5\degr$ and $1\degr$, but then it increases again and peaks in the annulus between $2\degr$ and $3\degr$ (see Fig. \ref{fig6}). 
The detected temperature asymmetry is much more extended than the visible M33 disk and seems to be present up to a galactocentric distance of about $3\degr$. 
This apparently odd behavior  indicates either a change in the emission geometry or a drastic change of the dominant emission mechanism, even if  the situation might be complicated by  contaminating effects, such as gas flowing into the M33 galaxy, and/or related to a past M33-M31 interaction (see, in particular, \citealt{bekki2008,wolf2013}).

Five possible emission mechanisms might explain the observed  temperature asymmetry: ($i$) free-free emission, ($ii$) synchrotron emission, ($iii$) anomalous microwave emission (AME) from dust grains, ($iv$) Sunyaev-Zel'dovich (SZ) effect, in particular the kinetic-SZ effect, and ($v$) cold gas clouds populating the
halo of M33. The absence of a substantial dependence of the observed temperature asymmetry on the considered {\it Planck} bands together with the fact that the hot and cold side are toward the expected directions indicates that it has to be modulated by the Doppler effect induced by the M33 spin. 

The dominating emission mechanism in the microwave regime is not clear as yet. Mechanisms  ($i$),  ($ii$), and  ($iv$) require the presence of a hot plasma with cosmic-ray electrons (CRE) around M33, however, and give rise to an emission spectrum  with a rather steep dependence on the frequency (see, e.g., \citealt{bennett2003,planck2014}). This plasma, expected to emit mainly in X-rays, is hard to detect since the emission measure scales with the square of the electron density, and when the galaxy surface brightness drops to the level of the X-ray background, mainly due to the Milky Way,  further detection is impossible. A hot gaseous halo (or corona) has been detected only for a few
spiral galaxies. In our Galaxy, \cite{gupta} probed the warm-hot phase (at a temperature $\simeq 10^6$ K) of the Galactic halo through the Chandra detection of $O_{VII}$ and $O_{VIII}$ absorption lines (with redshift $z=0$) toward known AGNs and inferred a warm halo extending up to $\simeq 100$ kpc and possibly containing about $10^{10}~M_{\odot}$ in gaseous form \citep{miller2013}. This gaseous halo probably rotates  with a velocity lower than the Local Standard of Rest velocity $v_{\rm LSR}$ \citep{Hodges-Kluck2016}.
For UGC12591, a hot halo has been detected that extends up to 110 kpc from the galactic center (\citealt{dai2012}). For  M33, although  a  hot-diffuse gas component is clearly visible in the XMM-Newton EPIC data up to about $30\arcmin$ (\citealt{pietsch2003}), it cannot be excluded that it extends farther out and, moreover, investigation of the CRE propagation length  shows that its value is even higher than that in M31 
(\citealt{berkhuijsen2013}). AME, option ($iii$), has been observed in various interstellar environments such as in the ISM \citep{miville2008} and in dark clouds \citep{watson2005}. Several processes can excite dust grain emission, provided they are present in galactic halo environments (see below): IR and radio photons, gas or dust interaction, formation of $H_2$ molecules on the dust grain surface, photoelectric emission (see, e.g., \citealt{draine1998}). Option ($v$) cannot be excluded at present because cold gas clouds (with or without a dust component) may populate the M33 halo, giving rise, if it were to rotate, to a certain temperature asymmetry through Eq. (\ref{eq1}). These clouds may be optically thin or thick to their own  sub-millimeter radiation,  depending on their mass and temperature. When the clouds are in virial equilibrium, the cloud optical thickness may be estimated through the relation 
 \begin{equation}
 \langle\tau\rangle\simeq  6\times 10^{-2} \left(\frac{T_c}{2.7~ {\rm K}} \right)^2  \left(\frac{M_c}{M_J} \right)^{-1},
 \end{equation}
where $M_c$ and   $T_c$ are the cloud mass and temperature, and $M_J$ is the mass of Jupiter. Large clouds are therefore expected to be optically thin, while only extremely low massive clouds (with $M_c\leq 0.1 M_J$) may be optically thick. The expected effect in the microwave bands is therefore very small. Cold clouds, such as those implied by BOOMERANG \citep{boomerang2010}, {\it Planck} \citep{planck2011}, and {\it Herschel} \citep{juvela2010,juvela2012}, also called{\it Herschel} cold clouds (HCC) and studied mainly toward the Small Magellanic Clouds (see \citealt{nieuwenhuizen2012} and references therein), might also be present in the M33 halo and  give rise to a temperature asymmetry in the {\it Planck} bands. We also mention that HI clouds have been discovered in the outer regions of M33  \citep{grossi2008,keenan2016}, and it has been suggested that they play a role in feeding  the star formation in the disk of M33. The observed gaseous features extend up to 38 kpc from the galaxy center,  and the HI mass beyond the M33 disk contains about $18\%$ of the estimated total HI mass \citep{putman2009}.
Until now, {\it Herschel} data of M33, within the project HERM33ES ({\it Herschel} M33 Extended Survey), have been analyzed and  reported in the literature only up to a galactocentric distance of about $0.5-0.6\degr$ and show a cold dust component with temperature $\simeq 14$ K between about 2 kpc and 6 kpc, and both SPIRE and PACS maps show a weak emission extending outside of 8 kpc \citep{kramer2010}.

For  the extension of the microwave emission toward M33 and its connection with some galactic population, we mention that in addition to the possible hot corona of CRE and plasma, there is  another component that has recently been discussed in the literature:  the RGB star halo population of M33, which was discovered by using the Pan-Andromeda Archeological Survey (PAndAS). This is found to have an exponential scale length of about $20$ kpc (about $1.4\degr$), as obtained  by  \cite{cockcroft2013}, and might account for the emission in the microwave bands through the mechanisms discussed above.  Even though RGB  stars cannot generate an asymmetry in the microwave temperature, a gas or
dust component (as that described above)  might be present in the outer region of the M33 galaxy, in association with the RGB stars. 

Galactic halos do contain the fossil record of galaxy assembly, and clarifying their composition is extremely important to understand the processes that led to the formation and subsequent evolution of these structures. The  degree to which galactic halos rotate with respect to the disks is also a relevant and extremely difficult question to answer (see, e.g., \citealt{courteau2011,deason2011}). All this indicates the need for more detailed, extended, and dedicated surveys.

\begin{acknowledgements}
We acknowledge the use of {\it Planck} data in the Legacy Archive for
Microwave Background Data Analysis (LAMBDA) and HEALPix
\citep{gorski2005} package. FDP, AAN, and GI acknowledge the support by the INFN
project TAsP, and PJ acknowledges support from the
Swiss National Science Foundation. 
AQ is grateful for hospitality to the DST Centre of Excellence in Mathematical \& Statistical Sciences of the University of the Witwatersrand, Johannesburg, South Africa.  The anonymous referee is acknowledged for the detailed comments.
\end{acknowledgements}


\end{document}